# Polarized Cold-neutron Reflectometry at JRR-3/MINE2 for the Development of Ultracold-neutron Spin Analyzers for a Neutron EDM Experiment at TRIUMF

Takashi Higuchi[1,2*], Hiroaki Akatsuka[3], Alexis Brossard[4], Derek Fujimoto[4], Pietro Giampa[4], Sean Hansen-Romu[5], Kichiji Hatanaka[2†], Masahiro Hino[1‡], Go Ichikawa[6], Sohei Imajo[2], Blair Jamieson[7], Shinsuke Kawasaki[8], Masaaki Kitaguchi[3,9], Russell Mammei[7], Ryohei Matsumiya[2,4], Kenji Mishima[6], Rüdiger Picker[4,10], Wolfgang Schreyer[11], Hirohiko M. Shimizu[3], Steve Sidhu[4,10], and Sean Vanbergen[4,12]

[1]*Institute for Integrated Radiation and Nuclear Science, Kyoto University, Kumotori, Osaka 590-0494, Japan*
[2]*Research Center for Nuclear Physics, Osaka University, Ibaraki, Osaka 567-0047, Japan*
[3]*Department of Physics, Nagoya University, Nagoya 464-8602, Japan*
[4]*TRIUMF, Vancouver, BC, Canada*
[5]*The University of Manitoba, Winnipeg, MB, Canada*
[6]*Institute of Materials Structure Science, KEK, Tokai, Ibaraki 319-1106, Japan*
[7]*The University of Winnipeg, Winnipeg, MB, Canada*
[8]*Institute of Particle and Nuclear Studies, KEK, Tsukuba, Ibaraki 305-0801, Japan*
[9]*Kobayashi-Maskawa Institute, Nagoya University, Nagoya 464-8602, Japan*
[10]*Simon Fraser University, Burnaby, BC, Canada*
[11]*Oak Ridge National Laboratory, Oak Ridge, TN, the United States of America*
[12]*The University of British Columbia, Vancouver, BC, Canada*

(Received September 30, 2023; revised July 16, 2024; accepted July 17, 2024; published online August 28, 2024)

The neutron electric dipole moment (EDM) is a sensitive probe for currently undiscovered sources of charge-parity symmetry violation. As part of the TRIUMF Ultracold Advanced Neutron (TUCAN) collaboration, we are developing spin analyzers for ultracold neutrons (UCNs) to be used for a next-generation experiment to measure the neutron EDM with unprecedented precision. Spin-state analysis of UCNs constitutes an essential part of the neutron EDM measurement sequence. Magnetized iron films used as spin filters of UCNs are crucial experimental components, whose performance directly influences the statistical sensitivity of the measurement. To test such iron film spin filters, we propose the use of polarized cold-neutron reflectometry, in addition to conventional UCN transmission experiments. The new method provides information on iron film samples complementary to the UCN tests and accelerates the development cycles. We developed a collaborative effort to produce iron film spin filters and test them with cold and ultracold neutrons available at JRR-3/MINE2 and J-PARC/MLF BL05. In this article, we review the methods of neutron EDM measurements, discuss the complementarity of this new approach to test UCN spin filters, provide an overview of our related activities, and present the first results of polarized cold-neutron reflectometry recently conducted at the MINE2 beamline.

## 1. Introduction

### 1.1 Background

The Standard Model (SM) of particle physics, although very successful, is known to be incomplete and is believed to be a low-energy limit of a more fundamental theory. One of the most compelling indications for the incompleteness of the SM is the dominance of matter over antimatter in the universe, commonly known as the baryon asymmetry problem. In 1967, Sakharov proposed a scenario to explain the baryon asymmetry based on a few conditions, including violation of charge-parity (CP) symmetry.[1] However, estimates based on SM CP violation fall short of explaining the observed asymmetry by many orders of magnitude. This has motivated the search for undiscovered sources of CP violation across various sectors of particle physics.

The neutron electric dipole moment (EDM) is crucial in this context. A nonzero EDM of a non-degenerate system, such as the neutron, would violate time-reversal symmetry. This is equivalent to CP violation based on the CPT theorem, which is fundamental to relativistic quantum field theories.[2,3] Consequently, the neutron EDM serves as a sensitive probe for CP violation. Starting from the first direct measurement by Smith, Purcell, and Ramsey,[4] experimental efforts to measure the neutron EDM have been pursued for decades, reaching the current best limit of $1.8 \times 10^{-26}\,e\mathrm{cm}$ (90% C.L.),[5] which sets stringent constraints on theories beyond the SM. At this time, multiple experiments worldwide are aiming for neutron EDM measurements with an order-of-magnitude improved sensitivity.[6] The theoretical prediction from the SM CP violation, however, lies on the order of $10^{-32}\,e\mathrm{cm}$.[7] Therefore, any finite neutron EDM observed in the next-generation experiments immediately implies a new source of CP violation.

### 1.2 Principle of a neutron EDM measurement

Experimentally, the neutron EDM is obtained by measuring the spin precession frequencies of neutrons in the presence of an electric field applied parallel or antiparallel to a magnetic field. The Hamiltonian of the neutron with magnetic moment $\boldsymbol{\mu}_n$ and EDM $\boldsymbol{d}_n$, under magnetic and electric fields $\boldsymbol{B}$ and $\boldsymbol{E}$ is represented by

$$H = -\boldsymbol{\mu}_n \cdot \boldsymbol{B} - \boldsymbol{d}_n \cdot \boldsymbol{E}. \quad (1)$$

Denoting the spin precession frequencies under configurations $\boldsymbol{B} \parallel \boldsymbol{E}$ (parallel) and $\boldsymbol{B} \parallel -\boldsymbol{E}$ (antiparallel) as $\omega_{\uparrow\uparrow}$ and $\omega_{\uparrow\downarrow}$ respectively, the EDM $d_n$ is extracted by

$$d_n = \frac{\hbar(\omega_{\uparrow\uparrow} - \omega_{\uparrow\downarrow})}{4|E|}. \quad (2)$$

©2024 The Author(s)

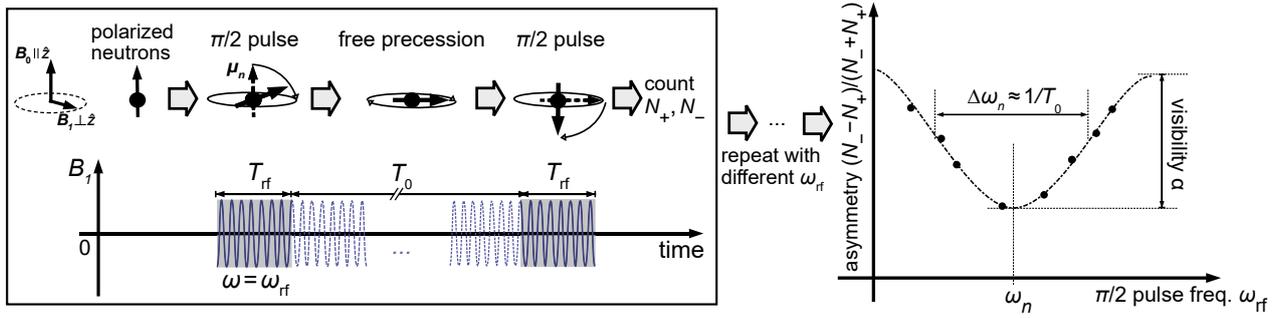

**Fig. 1.** (Color online) The principle of Ramsey's technique of separately oscillating fields. On the left, the sequence consisting of the first $\pi/2$ pulse with duration $T_\text{rf}$, the free precession time $T_0$, and the second $\pi/2$ pulse identical to the first one, is described. At the end of each sequence, the polarization of the neutrons is measured by counting the neutrons in each state. By repeating the sequence with different frequencies of the $\pi/2$ pulses $\omega_\text{rf}$, a fringe pattern on the right is obtained. The neutron spin-precession frequency $\omega_n$ is then determined by fitting the experimentally obtained fringe to the theoretical line shape described in Eq. (3).

Ramsey's technique of separately oscillating fields[8] is used to measure the spin-procession frequency in each configuration $\omega_n$ where $n \in \{\uparrow\uparrow, \uparrow\downarrow\}$. The measurement sequence used in this method is illustrated in Fig. 1. First, neutrons are polarized along the static magnetic field $\boldsymbol{B}_0$. A rotating magnetic field $\boldsymbol{B}_1$ with an angular frequency of $\omega_\text{rf}$ is applied perpendicular to $\boldsymbol{B}_0$ for duration $T_\text{rf}$ to rotate the spin by $\approx \pi/2$. Subsequently, the neutron spins precess freely for time $T_0$, and finally, the spin is rotated by another identical pulse phase-locked to the first. The final state polarization is measured by detecting neutrons in each spin state at the end of the cycle. From the measurement, the spin asymmetry $A = (N_- - N_+)/(N_- + N_+)$ is obtained, where $N_+$ and $N_-$ represent the neutron counts in the spin (+) and (−) states, respectively.

Repeating the sequence with different $\omega_\text{rf}$ results in an interference fringe of $A$, as shown in the figure. Near the resonance, i.e., $\omega_\text{rf} \approx \omega_n$, the fringe line-shape is approximated as

$$A(\omega_\text{rf}) = A_\text{off} - \alpha \cos\left(\frac{\omega_\text{rf} - \omega_n}{\Delta\omega_n}\right), \quad (3)$$

where $A_\text{off}$ and $\alpha$ are parameters representing the offset of the asymmetry and the fringe visibility respectively. The resonance width is expressed as $\Delta\omega_n \approx 1/T_0$ for $T_\text{rf} \ll T_0$. By fitting the fringe to Eq. (3), as shown in Fig. 1, the precession frequency $\omega_n$ is determined from the center of the resonance with an uncertainty of

$$\sigma(\omega_n) = \frac{\Delta\omega_n}{\alpha\sqrt{N}}, \quad (4)$$

where $N$ is the sum of the number of neutrons counted in all cycles.[9] Applying this to Eq. (2) assuming $\sigma(\omega_{\uparrow\uparrow}) = \sigma(\omega_{\uparrow\downarrow}) = \sigma(\omega_n)$, an estimate of the statistical sensitivity of the neutron EDM measurement is found to be

$$\sigma(d_n) = \frac{\hbar}{2\alpha|E|T_0\sqrt{N}}. \quad (5)$$

Since the 1970s, ultracold neutrons (UCNs) have been employed for neutron EDM measurements. This is because UCNs, which refer to neutrons with energies $\lesssim 300$ neV, can be stored in a material vessel whose surface has a high Fermi potential, often achieved via a thin coating. This substantially reduced the major systematic error due to a pseudomagnetic field proportional to $\boldsymbol{v} \times \boldsymbol{E}$,[10] and also contributed to the statistical sensitivity of the measurement [Eq. (5)] by extending the free precession time to the order of $T_0 \sim 100$ s. However, the challenge in the use of UCNs is the small number of available UCNs, which limits today's measurement sensitivity.

### 1.3 TRIUMF ultracold advanced neutron collaboration

In this context, we have formed the TRIUMF Ultracold Advanced Neutron (TUCAN) collaboration to build a high-intensity UCN source that will overcome the current statistical limitation and enable a neutron EDM measurement with a sensitivity of the order of $10^{-27}$ ecm. Our UCN production scheme is a combination of a spallation reaction driven by an accelerator beam and the so-called super-thermal UCN production by inelastic scattering of cold neutrons in superfluid helium.[11,12] The method was demonstrated using a prototype UCN source to produce the first UCNs at TRIUMF in 2017.[13] Currently, we are building a new upgraded UCN source that can be operated with a 40 µA proton beam from the TRIUMF cyclotron.[14] With the new UCN source, the UCN density in the nEDM precession chamber is expected to be $\approx 250$ UCN/cm$^3$ at the beginning of the measurement sequence, an improvement of two orders of magnitude as compared to the condition of the current best measurement.[5,15]

In parallel, essential subsystems of the neutron EDM spectrometer, including components for magnetic field control[16,17] and UCN handling,[18,19] are being developed. In this article, we focus on the development of UCN spin analyzers for the TUCAN EDM experiment.

### 1.4 Spin polarization and analysis of UCNs
### Principle and implementation in an nEDM measurement

A UCN spin analyzer is a crucial component of neutron EDM experiments based on Ramsey's technique. As described in Sect. 1.2, the analysis of neutron spin states is performed at the end of each measurement sequence by obtaining neutron counts $N_+$ and $N_-$ by the use of a UCN spin analyzer. The high performance of the spin analyzer is critical to the measurement, as it is required to achieve high visibility $\alpha$, which directly affects the statistical sensitivity of the EDM.

  ©2024 The Author(s)

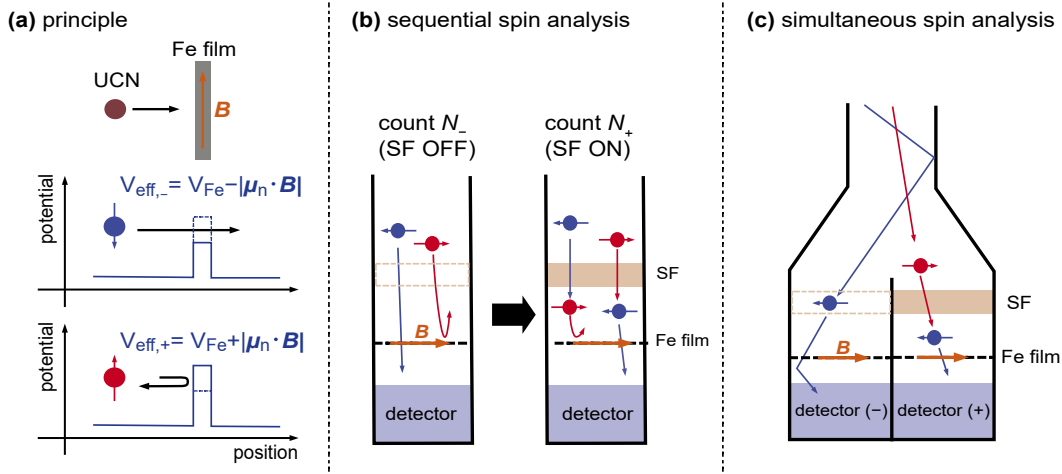

**Fig. 2.** (Color online) Principles and methods of UCN spin analysis with magnetized iron films. (a) Illustration of the effective potential experienced by UCNs in different spin states. Because the energies of UCNs are ≲300 neV, this potential creates a sufficient barrier that selectively transmits spin (−) UCNs. (b) Concept of sequential spin analysis by the use of an iron film spin filter, a detector, and a spin flipper. (c) Concept of simultaneous spin analysis with a set of two spin filers detectors and spin flippers.

Because of their extremely low energies, UCNs can be fully polarized by a magnetic field of a few tens of kilogauss. Thin iron films, which can be easily magnetized to ∼20 kG, are commonly utilized as spin filters to selectively transmit spin (−) UCNs. Upon passing through a magnetized iron film, a UCN experiences a spin-dependent effective potential[20]

$$V_{\text{eff},\pm} = V_{\text{Fe}} \pm |\mu_n \cdot B| \approx 209\,\text{neV} \pm 6\,\text{neV/kG} \cdot B, \quad (6)$$

where $V_{\text{Fe}} \approx 209$ neV denotes the Fermi potential of iron. As depicted in Fig. 2(a), the in-plane magnetic flux density $B \approx 20$ kG creates a sufficient potential difference between the two spin states, allowing only spin (−) UCNs to transmit through the film. Together with a spin flipper as in Fig. 2(b), such a spin filter can be used to count the number of UCNs in each spin state.[9] When a spin flipper placed above the magnetized iron film is turned off, spin (−) UCNs are counted, whereas spin (+) UCNs are stored above the film. When the spin flipper is turned on, the spin (+) UCNs are flipped to (−), and are then transmitted and counted. This allows UCNs in each spin state to be counted sequentially. To improve this further, the nEDM collaboration at PSI developed a two-arm analyzer, as shown in Fig. 2(c), which allows for the simultaneous analysis of both spin states by activating the spin flipper on one of the two arms. They reported an 18% increase in the EDM statistical sensitivity as compared to the sequential analysis of Fig. 2(b), owing to increased visibility and statistics.[21]

*UCN transmission measurement for testing spin filters*

In Fig. 3, a typical setup of a UCN transmission experiment to test the spin filters is sketched. It comprises two spin filters, two spin flippers, and a UCN detector. The UCNs are polarized by passing through the first film. By activating one of the spin flippers, the polarized UCNs flip their spins and are blocked at the second filter. Thus, the performance of the analyzer and the polarizer are evaluated from the reduction of the transmission by the spin-flipper operations. To interpret the results of such an experiment, a 2 × 2 matrix formalism is employed, where the UCN spin state is represented by a two-state spinor by choosing

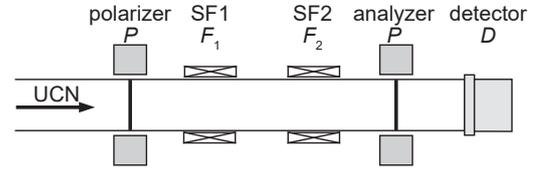

**Fig. 3.** Typical setup of a UCN transmission experiment to test the performance of UCN spin filters. Two iron-film spin filters are used as a polarizer and an analyzer (both expressed by $P$). Two spin flippers are inserted between the polarizer and the analyzer for spin manipulation ($F_1$ and $F_2$). The detector ($D$) is placed at the end.

$$|-\rangle = \begin{pmatrix} 1 \\ 0 \end{pmatrix}, \quad |+\rangle = \begin{pmatrix} 0 \\ 1 \end{pmatrix} \quad (7)$$

as the basis. The transfer matrices of the spin filters $P$, the detector $D$, and each of the spin flippers $F_1$, $F_2$ are expressed as

$$P = \begin{pmatrix} a_{11} & a_{12} \\ a_{21} & a_{22} \end{pmatrix}, \quad D = \begin{pmatrix} 1 & 1 \end{pmatrix},$$

$$F_1 = \begin{pmatrix} \epsilon_1 & 1-\epsilon_1 \\ 1-\epsilon_1 & \epsilon_1 \end{pmatrix}, \quad F_2 = \begin{pmatrix} \epsilon_2 & 1-\epsilon_2 \\ 1-\epsilon_2 & \epsilon_2 \end{pmatrix}. \quad (8)$$

In the ideal case, $a_{11} = 1$, $a_{12} = a_{21} = a_{22} = 0$, and $\epsilon_1 = \epsilon_2 = 0$. The UCN counts that can be obtained by different configurations of the two spin flippers are expressed as

$$N_{ij} = DP(F_2)^j(F_1)^i P \begin{pmatrix} 1 \\ 1 \end{pmatrix} \quad (i,j \in \{0,1\}), \quad (9)$$

where 0 and 1 of the index $i$ or $j$ correspond to the off and on of the spin flipper state, respectively. Equation (8) contains six unknown parameters, whereas only four experimental configurations are possible. Therefore, to evaluate the performance of the spin filter based on these experimental observables, an assumption must be made on the matrix elements of $P$. In previous studies, Egorov et al.[22] assumed $a_{12} = a_{21} = 0$, whereas Herdin et al.[23] used $a_{12} = a_{22} = 0$. These two assumptions for $P$ lead to different expressions that relate the spin filter performance as a UCN polarizer to the experimental observables $N_{ij}$.[24,25] Thus, there is some ambiguity as to which factor should be attributed as the cause of an imperfection in polarization, as each matrix element of $P$ represents a different mechanism of losing polarization. $a_{22} > 0$ could be caused by imperfect magnetization of the film, allowing a fraction of spin (+) UCNs to pass through the film, whereas nonzero $a_{12}$ or $a_{21}$ represents the spin flips of UCNs caused by traversing the film. It should also be noted that these factors have different significance depending on whether the spin filter is used as a polarizer or analyzer. When used as a spin analyzer, as shown in Figs. 2(b) and 2(c), nonzero $a_{21}$ does not affect the performance of the system as a spin analyzer, whereas nonzero $a_{12}$ or $a_{22}$ would result in misidentification of UCN spin states and could cause a reduction in the visibility of the Ramsey fringe, eventually worsening the EDM statistical sensitivity.

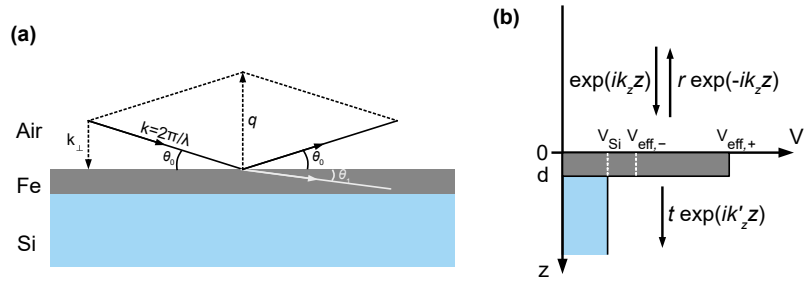

**Fig. 4.** (Color online) (a) Illustration of the beam paths in cold-neutron reflectometry. The incident angle $\theta_0$, the refractive angle into the iron layer $\theta_1$ and the momentum transfer $q$ are defined in the sketch. (b) One-dimensional potential used to derive the reflectivity model function. The thickness of the iron layer $d$ is one of the model parameters. The silicon layer, which is 2 mm thick, is treated as having an infinite thickness. The coefficients $r$ and $t$ are obtained by solving the one-dimensional Schrödinger equation. The reflectivity $R$ is related to the amplitude of the reflected wave $r$ by $R = |r|^2$.

**Table I.** The Fermi potential from the literature,[26] the critical momentum transfer $q_c$, and corresponding critical angle of total reflection $\theta_c$ for the 0.88 nm wavelength neutrons of the JRR-3/MINE2 beamline.

|  | Fermi potential (neV) | Critical momentum transfer $q_c$ (nm$^{-1}$) | Critical angle $\theta_c$ with $\lambda = 0.88$ nm (°) |
|---|---|---|---|
| Si | 54 | 0.102 | 0.410 |
| Ni | 245 | 0.217 | 0.873 |
| Magnetized Fe $V_{\text{eff},+}$ ($B = 20$ kG) | 329 | 0.252 | 1.011 |
| Magnetized Fe $V_{\text{eff},-}$ ($B = 20$ kG) | 89 | 0.131 | 0.526 |

## 2. Development of UCN Spin Analyzers for TUCAN

### 2.1 Strategies

In the following, we report recent work on the development of iron-film UCN spin filters for the TUCAN experiment based at the Institute for Integrated Radiation and Nuclear Science, Kyoto University (KURNS).

In previous studies, thin iron films deposited on aluminum foils[21,25] or silicon wafers[9,27] have been used. An advantage of using aluminum foil as a substrate is that it can be made as thin as a few tens of micrometers. Together with the small neutron absorption cross-section of aluminum, neutron transmission can be maximized. On the other hand, silicon wafers have a unique feature in that they can be prepared with polished mirror surfaces. Taking advantage of this property of silicon-substrate thin films, we propose the use of polarized cold-neutron reflectometry to characterize UCN spin filters. Cold neutrons are more accessible than UCNs and the experiments can be performed with higher statistics. If this new approach proves to provide reliable information on the performance of UCN spin filters, it is expected to greatly accelerate the development process. Furthermore, as described in detail in the next section, this method provides complementary information to the UCN transmission experiments.

With this motivation, we have started a joint effort to test iron film spin filters produced at KURNS by taking advantage of the capabilities of the neutron facilities at Material and Life Science Experimental Facility (MLF), J-PARC, and the Japan Research Reactor-3 (JRR-3). A unique pulsed UCN source at J-PARC/MLF BL05[28] allows UCN transmission tests of iron films on aluminum or silicon substrates. A long wavelength monochromatic beam from the JRR-3/C3-1-2 (MINE2) beamline[29] is suitable for cold-neutron reflectometry testing of iron films on silicon substrates.

### 2.2 Cold-neutron reflectometry for tests of UCN spin filters

Cold-neutron reflectometry is a versatile tool to study surfaces through the nuclear potential of the medium experienced by neutrons. Here, we apply it to single-layer iron films sputtered on a silicon wafer [Fig. 4(a)]. A reflectivity $R$ is characterized as a function of a momentum transfer upon reflection $q = 4\pi \sin\theta_0/\lambda$, with $\lambda$ being the wavelength of the neutron beam and $\theta_0$ the incident angle as defined in Fig. 4(a). As the reflection properties are determined by the momentum perpendicular to the surface, the properties of the film at UCN energies can be investigated using a cold neutron beam with a small incident angle. The critical momentum transfer $q_c$, below which neutrons are totally reflected, is related to the Fermi potential of the surface medium $V_F$ by

$$V_F = \frac{\hbar^2 q_c^2}{8 m_n}. \qquad (10)$$

In Table I, the values of the critical momentum transfer $q_c$ from the literature values of Fermi potential and the critical angle of total reflection $\theta_c$ with a wavelength of MINE2 beam (0.88 nm) are listed. By polarizing the incident beam, the reflection by the spin-dependent potential of Eq. (6) can be studied. The particular physical quantities that can be extracted from the measured reflectivity profiles are the reflectivity at the total reflection, the critical $q_c$ that corresponds to the potential which is experienced by the incident neutrons, and the thickness of the film $d$. By varying the magnetic field applied to the sample, these properties can be obtained as a function of the applied field. This provides important information as to how much of a magnetic field is sufficient for the iron film to be used as a high performance spin filter.

Recalling the discussion in Sect. 1.4, the reflectivity of the spin (+) neutrons $R_+$ is essentially

$$R_+ \approx 1 - (a_{12} + a_{22}) + \delta_{\text{abs}}, \quad (11)$$

with $\delta_{\text{abs}}$ representing the absorption probability, which is about 5% for UCN with a 50 nm wavelength. An iron film with a nearly perfect reflectivity is expected to have a high performance as a UCN spin analyzer. Given the above-discussed ambiguity in the interpretation of UCN transmission experiments, this method provides a complementary way to characterize the iron film spin filters.

### 2.3 Sample preparation and magnetic characterization

The ion beam sputtering facility at KURNS[30] is used to prepare samples of iron films. The sputtering stage has a diameter of 500 mm, a size sufficient to produce samples for large aperture UCN experiments. For samples tested at MINE2 in Sect. 3, a single layer of 97 nm thick iron (purity of 99.995%) is sputtered on the mirror surface of 2 mm thick silicon wafers.

The magnetization curves are measured with a vibrating sample magnetometer by cutting out a $\approx 1 \times 1 \text{ cm}^2$ piece from a witness sample simultaneously sputtered with the main samples. In Fig. 5, examples of the magnetization curves are shown for simultaneously sputtered 97-nm-thick iron films on a 0.38-mm-thick silicon wafer (a) and a 25-μm-thick aluminum foil (b). It can be seen that the iron film on the silicon substrate is magnetized with a smaller magnetic field compared to the aluminum substrate. The magnetization curve of the silicon substrate sample is fully saturated with about 50 Oe of the applied field, while about 200 Oe is required for the aluminum substrate sample. This is due to the strain in the iron film caused by the deformation of the aluminum foils, which changes the magnetic susceptibility of the iron film through the inverse magnetostrictive effect.[31]

### 2.4 Recent tests with a pulsed UCN source at J-PARC

A Doppler-shifter type UCN source at J-PARC/MLF BL05 produces UCNs by reflecting very cold neutrons with a velocity of 136 m/s off a backward-moving $m = 10$ supermirror mounted on a rotating arm.[28] Choosing the speed of the mirror to be half that of the incident neutrons, the neutrons are decelerated to near zero velocity and are delivered as pulsed UCNs at a repetition rate of 8.33 Hz.

A test setup of UCN spin filters with this source is shown in Fig. 6(a), which is essentially equivalent to that of Fig. 3. The same magnetic field is applied to the upstream and downstream iron films by the two dipole magnets. A set of two spin flippers consists of steel yokes that produce a gradient static magnetic field and solenoid coils that generate RF magnetic fields for adiabatic spin flipping.[32] A glass tube coated with sputtered NiMo is used as a UCN guide for the section with the spin flippers to allow RF field penetration.

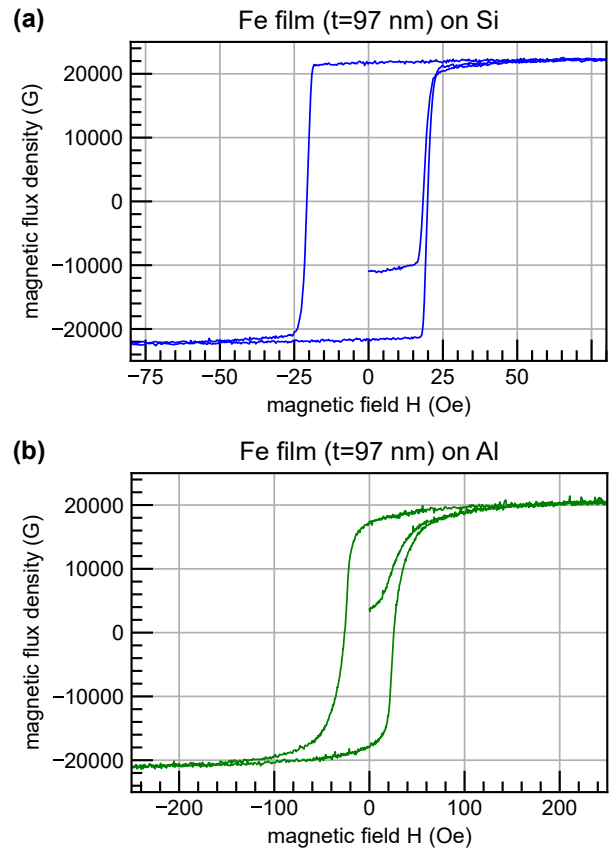

**Fig. 5.** (Color online) The magnetization curves of silicon-substrate (a) and aluminum-substrate iron films (b) obtained by vibrating sample magnetometry.

The energies of pulsed UCNs can be resolved by their time of flight. Examples of the UCN spectra are displayed in Fig. 6(b) for an experiment with silicon-substrate iron films under a 120 Oe applied magnetic field. A clear reduction of the transmitted UCNs is observed in the spin-flipper conditions $N_{01}$ and $N_{10}$, as compared to $N_{00}$ and $N_{11}$ [see Eq. (9)] for wavelengths >50 nm, corresponding to energies of <329 neV.

Systematic studies to understand the background neutron events are underway so that they can be properly accounted for in the data evaluation. The final polarization results of this experiment will be reported in a future publication.

## 3. Polarized Cold-Neutron Reflectometry at MINE2

### 3.1 Setups

In December 2022, we conducted the first experiment of polarized cold-neutron reflectometry at JRR-3/MINE2. The setup of Fig. 7 was constructed at the MINE2 beamline, which provided a 0.88 nm monochromatic neutron beam with a resolution of $\Delta\lambda/\lambda \sim 2.7\%$ at the full width at half maximum.[29] The setup consisted mainly of polarizing and analyzing mirrors, a resonant spin flipper (SF), a Helmholtz coil (H) to apply a magnetic field to a sample, and a $^3$He neutron detector (D). A rectangular guide coil covered the central part of the setup and provided a guide magnetic field of 7 Oe. M1 and M2 in Fig. 7 represent magnetic mirror holders equipped with an array of permanent magnets that

J. Phys. Soc. Jpn. **93**, 091009 (2024) Special Topics T. Higuchi et al.

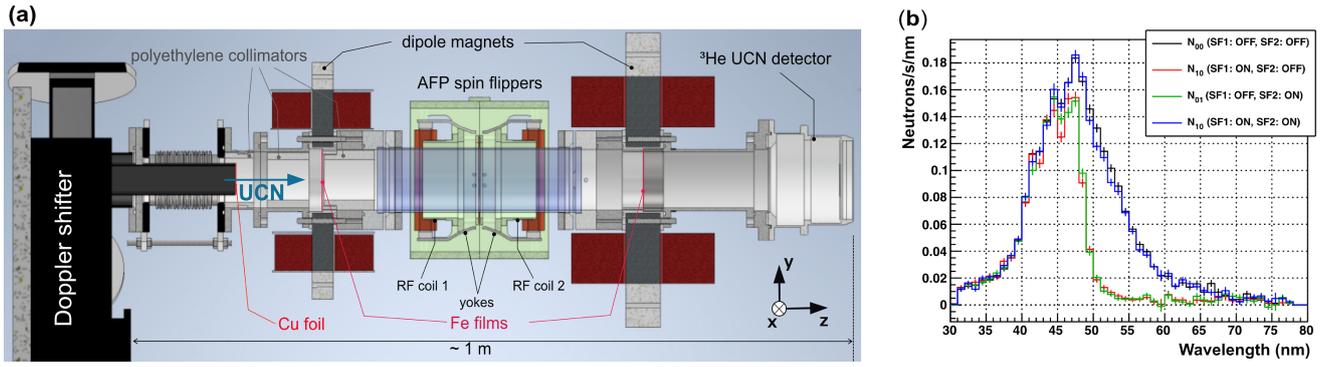

**Fig. 6.** (Color online) Setup for UCN transmission experiments with a pulsed UCN source at J-PARC/MLF BL05. Pulsed UCNs are produced by the Doppler-shifter type UCN source, and introduced into the setup. The two spin filters mounted in the dipole magnets and two spin flippers are used to polarize incoming UCNs and manipulate their spin states, as discussed in Sect. 1.4. (b) The UCN wavelength spectra from transmission experiments with iron film samples on silicon substrates under a magnetic field of 120 Oe. Data acquired for the four spin flipper conditions are presented.

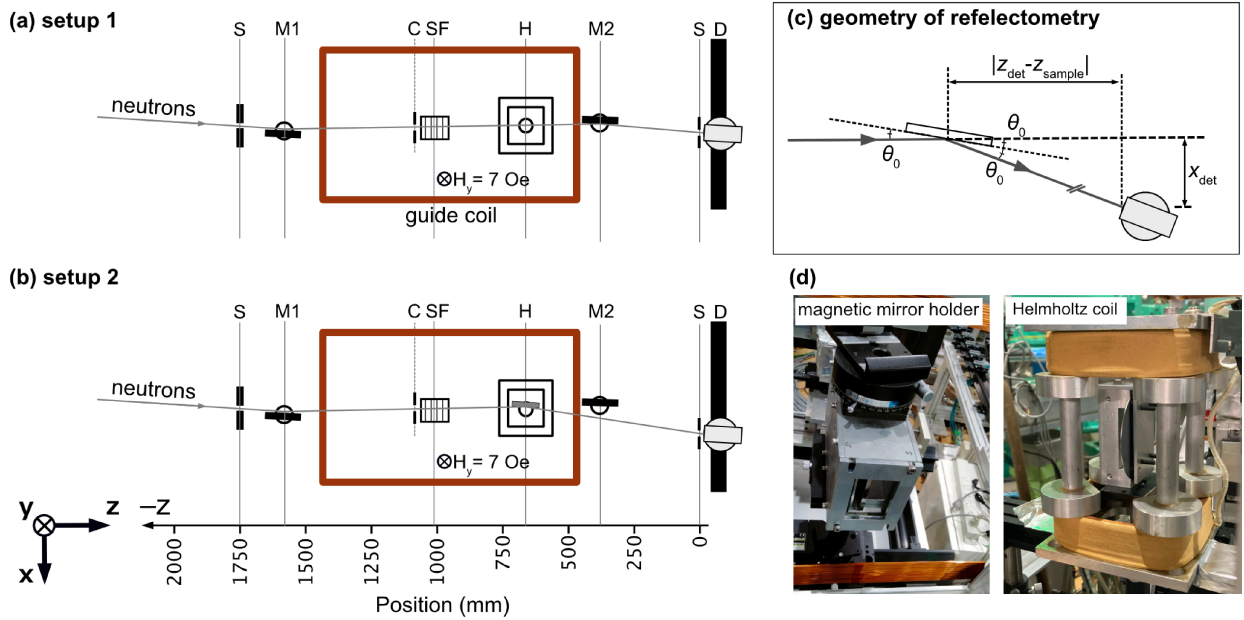

**Fig. 7.** (Color online) The setups of the polarized cold-neutron reflectometry conducted at the JRR-3/MINE2 beamline. S: slits, M1, M2: magnetic mirror holders, C: collimator, H: Helmholtz coil, D: detector. (a) Setup 1 for polarizer characterization and sample reflectivity measurement with a magnetic field of 300 Oe. (b) Setup 2 for sample reflectivity measurement up to 55 Oe of the applied field. (c) Zoomed-in view around the sample or the analyzing magnetic bilayer mirror to show the geometrical relations of the incident and reflected beams, and the detector position and orientation. During the angle scan, the detector position and orientation is incrementally moved to detect the reflected beam, which forms twice the angle of the incident angle with the incident beam. (d) Photographs of one of the magnetic mirror holders and the Helmholtz coil.

applied 500 Oe (M1) or 300 Oe (M2) to the reflectometry element in the holder.

Setup 1 in Fig. 7(a) was used to characterize the beam polarization in the polarizer-analyzer configuration with the magnetic multilayer mirrors installed in both M1 and M2. Setup 2 was used to measure the reflectivity of the iron film sample under magnetic fields up to 55 Oe using the Helmholtz coil. To apply a higher field, we returned to setup 1, replaced the analyzing mirror in the magnetic mirror holder M2 with the iron film sample, and measured the sample reflectivity with an applied field of 300 Oe.

The sample/mirror holders and the detector were mounted on rotational and $x$-axis translational stages. During a scan of the incident angle $\theta_0$, the detector was moved to the anticipated position and orientation where the reflected beam forms an angle of $2\theta_0$ with the incident beam [Fig. 6(c)]. The $x$ coordinate of the detector $x_{det}$ is related to the incident angle $\theta_0$ by

$$2\theta_0 = \tan^{-1}\left(\left|\frac{x_{det}}{z_{det} - z_{sample}}\right|\right). \tag{12}$$

The flight distance $|z_{det} - z_{sample}|$ was determined by measuring the distance between the downstream side of the detector and the center of the rotational stage of the sample.

### 3.2 Characterization and optimization of the polarizer

A Fe/SiGe multilayer mirror was used as a polarizer.[33] The composition of the SiGe layer was designed such that the Fermi potential of this layer is equal to the $V_-$ of the magnetically saturated iron, i.e., ≈89 neV. This creates a grating that is only experienced by spin (+) neutrons. For the ones used in this experiment, the effective iron layer interval was about 14 nm, resulting in about 1.8° of the Bragg angle.

Figure 8 shows the results of characterization measurements performed with setup 1. One multilayer mirror was installed in M1 as a polarizer and the other in M2 as an analyzer. The polarizer was fixed to the incident angle of 1.78°, and the incident angle on the analyzer mirror was scanned. The intensity of the reflected beam was obtained as a function of the incident angle for the spin flipper off ($N_0$) and on ($N_1$) [Fig. 8(a)]. In Fig. 8(b), the beam polarization is estimated by the spin-flip asymmetry of the counts: $p_2 = (N_0 - N_1)/(N_0 + N_1)$. From these results, the polarizer was found to be most efficient at an incident angle of 1.78°, resulting in 95% asymmetry. This spin-flip asymmetry must be distinguished from the polarization power of the polarizer itself. If we denote the beam polarization power directly after the polarizer as $p_0$, that after the analyzer as $p_1$, and the spin-flipper efficiency as $1 - \epsilon_0$ ($0 \leq p_0, p_1, \epsilon_0 \leq 1$), they are related to $p_2$ by

$$p_2 = p_0(1 - \epsilon_0)p_1. \tag{13}$$

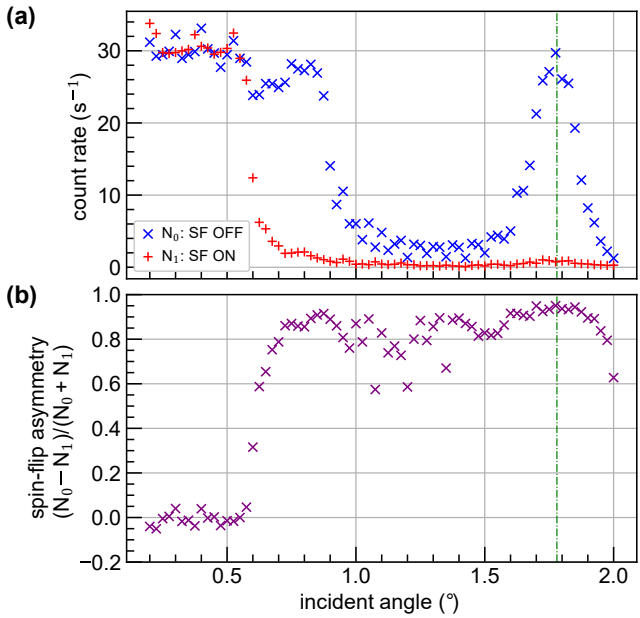

**Fig. 8.** (Color online) Characterization of magnetic multilayer mirror with setup 1 [Fig. 7(a)]. (a) The count rate of the reflected neutron beam as a function of incident angle on the analyzing mirror with the spin flipper off and on. (b) Polarization of the system calculated by the count rates in (a) with the spin flipper off ($N_0$) and on ($N_1$). The angle of 1.78° employed as the incident angle of the polarizer of the reflectometry measurement is indicated by the green dashed line, at which the maximum polarization in the reflected beam is obtained.

The maximum flipping asymmetry in Fig. 8(b) is about $p_2 = 0.95$ at 1.78°. To further improve the spin-flipper efficiency, a collimator is installed before the spin flipper [Figs. 7(a) and 7(b) "C"]. With this setup and the Helmholtz coil at 55 Oe, a polarization of $p_2 = 0.96(1)$ was achieved.

Previously, we attempted a similar reflectometry measurement using a pulsed cold-neutron beam at the low-divergence branch of J-PARC/MLF BL05.[34] In this case, the incident angle was fixed, and the $q$ scan was obtained by different wavelengths of neutrons resolved by their time of flight. The main challenge in this case was the wavelength dependence of the polarizing power of the polarizer and the efficiency of the spin flipper. In comparison, having a constant high polarization over the whole range of the $q$ scan is a great advantage of the monochromatic, long-wavelength neutron beam of MINE2.

**Table II.** Measurement parameters and results of magnetic field measurements by a Gauss meter for each measurement condition. The current applied to the Helmholtz coil, the magnetic fields measured at the center of the Helmholtz coil and the downstream end of the spin flipper, the frequency, and the amplitude of the spin flipper are listed. The amplitude of the spin flipper is set on the function generator before being fed to an amplifier.

| $I$ (A) | $H_{\text{sample}}$ (Oe) | $H_{\text{SF}}$ (Oe) | SF frequency (kHz) | SF amplitude (mV$_{\text{pp}}$) |
| --- | --- | --- | --- | --- |
| 0.00 | 6.9 | 7.0 | 20.8 | 55 |
| 0.80 | 17.0 | 6.8 | 20.6 | 53 |
| 1.45 | 25.0 | 6.6 | 20.5 | 55 |
| 2.00 | 32.0 | 6.2 | 20.3 | 53 |
| 4.00 | 55.0 | 5.5 | 19.8 | 52 |
| 0.00 | 300 | 7.0 | 20.8 | 55 |

### 3.3 Measurement of a silicon-substrate iron film sample

After achieving sufficient beam polarization, we performed reflectivity measurements of an iron film sample. The sample was an approximately 90-nm-thick iron film sputtered on a 2-mm-thick silicon wafer.

The reflectivity of the sample was measured at magnetic fields of 17, 25, 35, 55 (with setup 2), and 300 Oe (with setup 1). In setup 2, a reverse magnetic field was applied prior to the measurement to map defined points in the hysteretic magnetization curve. The field is then scanned as $-41 \rightarrow 17 \rightarrow 25 \rightarrow 35 \rightarrow 55$ Oe. The changes in the stray field from the Helmholtz coil were found to significantly alter the static magnetic field of the spin flipper. Therefore, the frequency and the amplitude of the spin flipper were re-optimized for each measurement condition. The magnetic fields measured by a Gauss meter and the spin-flipper parameters used for each condition are listed in Table II.

For each magnetic field condition, reflectivity is obtained as a function of the incident angle with the spin flipper off and on. Examples are shown for 55 and 300 Oe in Fig. 9.

*Fit model*

A model was developed as follows, to fit the obtained reflectivity profiles. First, the theoretical reflectivity of a single-layer film with a potential configuration in Fig. 4(b) is derived. The parameters $q_{1,\pm}$ and $q_2$ are defined as

$$q_{1,\pm} = \sqrt{\frac{8m_n V_{\text{eff},\pm}}{\hbar^2}}, \quad q_2 = \sqrt{\frac{8m_n V_{\text{Si}}}{\hbar^2}}. \tag{14}$$

Here, $q_{1,+}$ is the critical momentum transfer experienced by spin (+) neutrons, and $q_{1,-}$ that of the spin (−) neutrons. We assume an iron layer thickness of $d$ and an infinite thickness of silicon. By solving the one-dimensional Schrödinger equation with this boundary condition, the reflectivity is obtained from the amplitude of the reflected wave to be

$$R(q|q_{1,\pm}, q_2, d) = \left| \frac{r_{10} - r_{12} \exp(i\bar{q}_{1,\pm}d)}{1 - r_{12}r_{10}\exp(i\bar{q}_{1,\pm}d)} \right|^2 \tag{15}$$

with

$$\bar{q}_{1,\pm} \equiv \sqrt{q^2 - q_{1,\pm}^2}, \quad \bar{q}_2 \equiv \sqrt{q^2 - q_2^2} \tag{16}$$

and

$$r_{10} \equiv \frac{q - \bar{q}_{1,\pm}}{q + \bar{q}_{1,\pm}}, \quad r_{12} \equiv \frac{\bar{q}_2 - \bar{q}_{1,\pm}}{\bar{q}_2 + \bar{q}_{1,\pm}}. \tag{17}$$

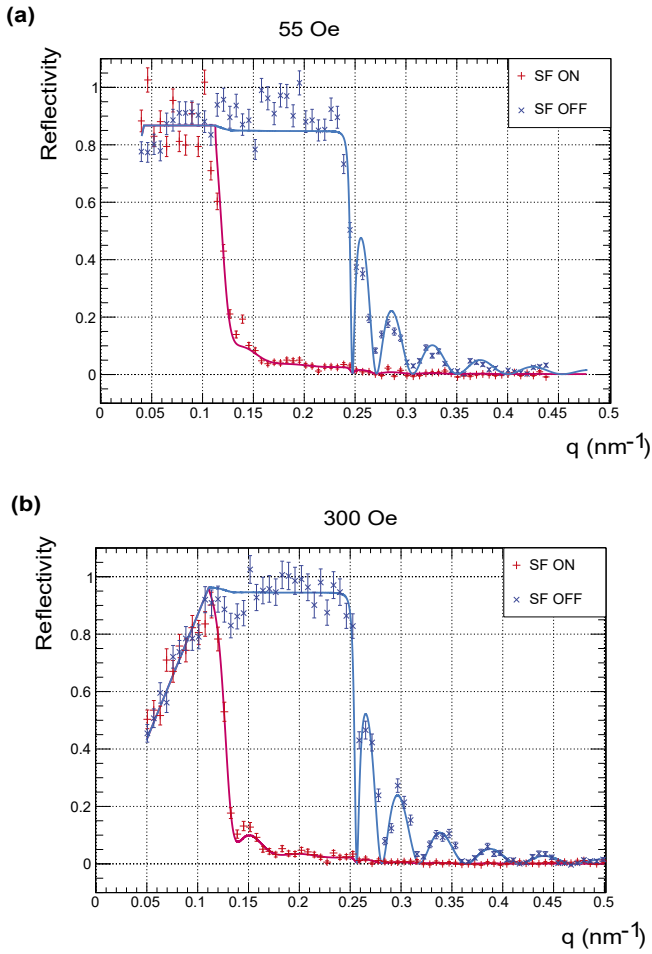

**Fig. 9.** (Color online) The reflectivity profiles obtained by the measurement. The reflectivity is plotted as a function of the momentum transfer $q$. Data with the spin flipper off and on, obtained for the 55 Oe (a) and 300 Oe (b) are displayed. The fit functions of Eqs. (20) and (21) are shown in the solid lines.

Based on this, three systematic effects are implemented in the fit model. First, if the film is not fully magnetized, a fraction of the neutrons will not experience the magnetic domains in the film aligned to the polarization axis, reducing the maximum reflectivity from 1.[35] Defining the maximum reflectivity as $\rho$, independent of the neutron spin states, this replaces $R(q|q_{1,\pm}, q_2, d)$ with $\rho R(q|q_{1,\pm}, q_2, d)$.

The second effect is the imperfect polarization of the incident beam. The asymmetry of spin (+) and (−) incident on the sample is a product of the polarization power of the polarizer and the spin-flipper efficiency: $p_0(1 - \epsilon_0)$ in the notation of Eq. (13). Redefining this as $p_3 = p_0(1 - \epsilon_0)$, the reflectivity with the spin flipper off ($R_0$) and on ($R_1$) are related to $R(q|q_{1,\pm}, q_2, d)$ by

$$R_0 = \frac{1+p_3}{2} R(q|q_{1,+}, q_2, d) + \frac{1-p_3}{2} R(q|q_{1,-}, q_2, d),$$
$$R_1 = \frac{1-p_3}{2} R(q|q_{1,+}, q_2, d) + \frac{1+p_3}{2} R(q|q_{1,-}, q_2, d). \tag{18}$$

The last effect to be considered is a geometric factor in the reflectometry. This is particularly evident in the data at 300 Oe obtained with setup 1 [Fig. 9(b)]. For a small incident angle, the beam footprint increases following $1/\sin(\theta_0)$. If the beam footprint exceeds the size of the sample, this reduces the counted neutrons and leads to an apparent decrease in the reflectivity. Defining the effective threshold in the unit of momentum transfer $q_{th}$ below which this loss occurs, this scaling factor $\eta(q)$ is expressed as

$$\eta(q|q_{th}) = \begin{cases} q/q_{th} & (q < q_{th}) \\ 1 & (q \geq q_{th}). \end{cases} \tag{19}$$

Putting these together, the reflectivity profiles with the spin flipper off $R'_0$ and on $R'_1$ are modeled by the following expressions:

$$R'_0(q|q_{1,+}, q_{1,-}, q_2, q_{th}, d, p_3, \rho)$$
$$= \rho \eta(q|q_{th}) \left[ \frac{1+p_3}{2} R(q|q_{1,+}, q_2, d) \right.$$
$$\left. + \frac{1-p_3}{2} R(q|q_{1,-}, q_2, d) \right], \tag{20}$$

$$R'_1(q|q_{1,+}, q_{1,-}, q_2, q_{th}, d, p_3, \rho)$$
$$= \rho \eta(q|q_{th}) \left[ \frac{1-p_3}{2} R(q|q_{1,+}, q_2, d) \right.$$
$$\left. + \frac{1+p_3}{2} R(q|q_{1,-}, q_2, d) \right]. \tag{21}$$

For each magnetic field condition, data of the reflectivity profiles with the spin flipper on and off were simultaneously fitted to the model to determine the seven unknown parameters. The fit functions for the magnetic field at 55 and 300 Oe are displayed in Fig. 9.

*Results*

For each magnetic field condition, we performed a simultaneous least-squares fit of the model [Eqs. (20) and (21)] to the data with the spin flipper on and off. The results obtained by a fit routine based on the MIGRAD algorithm[36] are listed for each dataset in Table III. The goodness of fit indicated by $\chi^2$/ndf is larger than ideal; 5.4–6.7 for the data with setup 1, and 2.7 for the data with setup 2.

From Table III, it can be seen that there is an increase in reflectivity between 55 and 300 Oe from 0.866(5) to 0.963(9). Even considering the goodness of fit, this is a significant change. The trend of higher reflectivity in the 300 Oe data compared to the 55 Oe data can also be seen in the reflectivity profiles in Fig. 9. This implies that the magnetic field of 55 Oe is not sufficient to fully magnetize the iron film to be used as an efficient spin analyzer, and the critical point of this rise in performance is somewhere between 55 and 300 Oe. It is also interesting to compare these results with the magnetization curve of Fig. 5(a), where from macroscopic observation, about 50 Oe of magnetic field seems to be sufficient for saturation.

At 17 Oe, the sample was still magnetized to $y < 0$ orientation, resulting in the swapped values between $q_{1,+}$ and $q_{1,-}$ as compared to the other magnetic field configurations.

Different values of the polarization in each condition are the consequence of the stray fields of the Helmholtz coil affecting the spin flipper. This then required re-optimizing the spin flipper in each condition, leading to slightly different spin-flipper tunes at each condition. To improve this aspect, it is planned to replace the Helmholtz coil with an electromagnet whose yokes will reduce stray fields. This approach enables scanning through a range of magnetic fields without changing the configuration of the spin flipper. It is also planned to replace the $x$ stage scan setup with a dedicated $\theta$–$2\theta$ stage consisting of two rotational stages with a common rotation axis: one for setting the incident angle on the sample and the other for moving the detector to the right position. This will allow for a more reliable determination of the incident angle to the sample.

**Table III.** Results of the reflectivity measurement at different magnetic fields. For each magnetic field condition, the reflectivity profiles with the spin flipper off and on are simultaneously fitted by the functions Eqs. (20) and (21) to determine the parameters in the table, including the critical momentum transfers, the maximum reflectivity $\rho$, and the polarization $p_3$.

| $H$ (Oe) | $q_{1,+}$ (nm$^{-1}$) | $q_{1,-}$ (nm$^{-1}$) | $q_2$ (nm$^{-1}$) | $q_{\text{th}}$ (nm$^{-1}$) | $d$ (nm) | $\rho$ | $p_3$ | $\chi^2/\text{ndf}$ |
|---|---|---|---|---|---|---|---|---|
| 17  | 0.1286(3) | 0.2475(1) | 0.1228(5) | 0.062(1) | 97.1(2) | 0.853(6) | 0.956(2) | 739/123 |
| 25  | 0.2472(1) | 0.1251(4) | 0.1214(7) | 0.062(1) | 96.6(2) | 0.881(6) | 0.905(4) | 709/123 |
| 32  | 0.2477(1) | 0.1264(3) | 0.1221(7) | 0.060(1) | 98.0(6) | 0.870(6) | 0.937(4) | 667/123 |
| 55  | 0.2476(1) | 0.1255(3) | 0.1219(6) | 0.052(2) | 97.0(2) | 0.866(5) | 0.955(3) | 825/123 |
| 300 | 0.2477(2) | 0.1214(5) | 0.111(3)  | 0.111(2) | 94.2(3) | 0.963(9) | 0.962(4) | 373/139 |

## 4. Conclusions

As part of the effort to develop UCN spin analyzers for the next generation neutron EDM experiment at TRIUMF, we conducted polarized cold-neutron reflectometry of a single-layer silicon-substrate iron film at MINE2. A polarized monochromatic cold-neutron beam with more than 96% polarization was produced, and the reflectometry measurement was performed by using the existing infrastructure of the beamline.

Although the experiment is still planned to be refined and upgraded, these first results already demonstrate the effectiveness of this method for the characterization of UCN spin filters, and the capability of the MINE2 beamline for such development work with polarized neutrons.

**Acknowledgements** We thank K. Ohno and the Research Fabrication Support Division of the Osaka University Core Facility Center for their engineering support of the UCN experiment, and C. Marshall and T. Lightbody (TRIUMF) for their valuable advice on apparatus design. We thank the assistance of D. Georgescu (the University of Edinburgh) in the UCN apparatus design, and I. Press (the University of Winnipeg) for experimental assembly, and J. Sato (Nagoya University) for an important remark on the neutron reflectivity model. We also thank H. Tsukahara (Osaka University & Tohoku University) and T. Hawai (CROSS) for valuable discussions at an early stage of this project.

The neutron experiments at the Materials and Life Science Experimental Facility of J-PARC were performed under a user program (Proposal Nos. 2021B0272, 2022B0329) and S-type project of KEK (Proposal No. 2019S03). The neutron experiment at JRR-3 was carried out by the JRR-3 general user program managed by the Institute for Solid State Physics, the University of Tokyo (Proposal No. 22589). The iron film fabrication work has been carried out under the visiting researcher's program of the Institute for Integrated Radiation and Nuclear Science, Kyoto University (Proposal Nos. R3137, R4099).

This work is financially supported by JSPS KAKENHI (Grant Nos. 18H05230, 19K23442, 20KK0069, 20K14487, and 22H01236), JSPS Bilateral Program (Grant No. JSPSBP120239940), JST FOREST Program (Grant No. JPMJFR2237), the Natural Sciences and Engineering Research Council of Canada (NSERC) SAPPJ-2023-00029, the Canada Foundation for Innovation, the Canada Research Chairs Program, International Joint Research Promotion Program of Osaka University, RCNP COREnet, the Yamada Science Foundation and the Murata Science Foundation.

---

*higuchi.takashi.8k@kyoto-u.ac.jp
†Deceased
‡hino.masahiro.2x@kyoto-u.ac.jp